\documentstyle[12pt,rotate,epsf]{article}
\def\kpc{\,{\rm kpc}}

\def\cmm2{{\,\rm cm^{-2}}}
\def\cm2{{\,{\rm cm}^2}}
\def\cmm3{{\,{\rm cm}^{-3}}}
\def\gcmm3{{\,{\rm g\,cm^{-3}}}}

\def\la{\mathrel{\mathpalette\fun <}}
\def\ga{\mathrel{\mathpalette\fun >}}
\def\fun#1#2{\lower3.6pt\vbox{\baselineskip0pt\lineskip.9pt
  \ialign{$\mathsurround=0pt#1\hfil##\hfil$\crcr#2\crcr\sim\crcr}}}

\begin{document}
\pagestyle{empty}
\begin{center}
\bigskip

\rightline{FERMILAB--Pub--96/022-A}
\rightline{astro-ph/9601168}

\vspace{.2in}
{\Large \bf MACHOs:  THE PLOT THICKENS}
\bigskip

\vspace{.2in}
Michael S. Turner,$^{1,2,3}$ Evalyn I. Gates,$^{2,3}$ and
Geza Gyuk$^1$\\

\vspace{.2in}
{\it $^1$Department of Physics \\
Enrico Fermi Institute, The University of Chicago, Chicago, IL~~60637-1433}\\

\vspace{0.1in}
{\it $^2$NASA/Fermilab Astrophysics Center\\
Fermi National Accelerator Laboratory, Batavia, IL~~60510-0500}\\

\vspace{0.1in}
{\it $^3$Department of Astronomy \& Astrophysics\\
The University of Chicago, Chicago, IL~~60637-1433}\\

\end{center}

\vspace{.3in}
\centerline{\bf ABSTRACT}
\bigskip
We discuss the implications of the recent
upward revision of the LMC microlensing rate by the
MACHO Collaboration.   We conclude:  (i)  A
good case for the existence of baryonic dark matter
in the halo has been made; (ii) The case for the existence
of cold dark matter is unaffected and still compelling;
and (iii) The Galactic
halo is still an excellent place to search for cold dark
matter particles (e.g., axions or neutralinos).

\newpage
\pagestyle{plain}
\setcounter{page}{1}
\newpage

\section{Perspective}

It is now a little more than two years since the MACHO
and EROS Collaborations announced their first microlensing results:
one LMC event for MACHO and two LMC events for EROS \cite{nature}.
At about the same time OGLE also reported
microlensing events toward the bulge, at a much higher rate than
expected rate \cite{ogle}.  Paczynski's bold idea had
become reality.

The initial response to the LMC results was swift and not unexpected:
The dark matter problem has been solved -- it was baryons
after all.  A dark cloud of gloom engulfed the
particle dark matter searches.  After a while,
sanity returned:  The numbers were small and the uncertainties -- both
Poisson and galactic modelling -- were large \cite{meaning}.

About a year ago MACHO \cite{machoprl} and EROS \cite{eroslmc}
presented detailed analyses of a big chunk of data (for MACHO
the first year's data with three events toward the LMC; for
EROS the same two events).  Both groups concluded that the
halo MACHO fraction was small, around 20\%.  Our independent
analysis \cite{prl,prd} of the LMC and bulge data
based upon extensive galactic modelling indicated the same
(likelihood function peaked around a halo MACHO fraction of
15\%.)  WIMP hunters were very  happy.  However,
as we and others (David Bennett especially) emphasized the
uncertainties were still large and it was too early to rule
out an all-MACHO halo.  Moreover, others had suggested
radical ideas for an all baryonic halo, e.g., gas clouds \cite{jetzer}.

The OGLE \cite{ogle} -- and later the MACHO \cite{machobulge}
-- higher-than-expected bulge results provided further evidence
that the bulge is actually a heavy bar, a possibility not considered in
the original estimates.  This has important consequences
for interpreting the LMC events.   A heavy bar
precludes a light halo, which, for a given LMC microlensing
rate, pushes down the halo MACHO fraction \cite{prl,prd}.

The MACHO Collaboration has now analyzed two
years of LMC data and has 8 or 7 or 6 events, depending upon
selection criteria.  Two year-one
events have now been rejected (one turned out to be a repeater).
The halo MACHO fraction is now quoted as 50\% \cite{nyt}.
The dark cloud threatens WIMP chasers again -- 50\% is
dangerous close to 100\%.

What happened?  The nearly 100 bulge events seen by the
MACHO Collaboration allowed a ``tune up''  of their event-recognition
software, and they have another full year of data.  They are able to
detect longer duration events -- and reject
repeaters -- and the statistics are better.

Beware!  The number of events is still small and galactic uncertainties
are still large.  More long duration events may be found;
some candidates may be rejected, and the software
may improve again -- leading to higher or lower
rates.  It is not time for WIMP hunters to panic.

\begin{figure}[htb]
\epsfxsize=7.5cm
\center\leavevmode\rotate[r]{\epsfbox{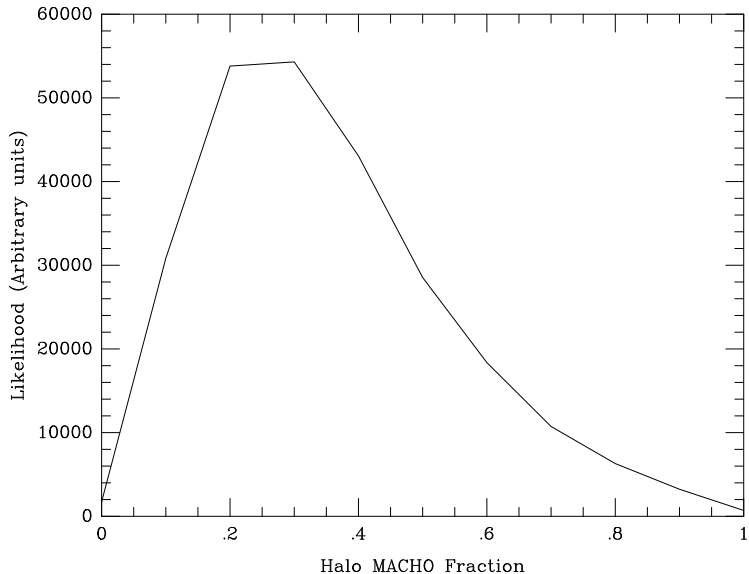}}
\caption{Monte-Carlo generated likelihood function for the
halo MACHO fraction based upon the upward revision of the
LMC optical depth by the MACHO collaboration \protect\cite{nyt,dbatucb}
and extensive galactic modelling (described in Refs.~\protect\cite{prl,prd}).}
\label{fig:1}
\end{figure}

Based upon the seven MACHO LMC events, the microlensing rate
toward the bulge and extensive modelling of the Galaxy, we find
a likelihood function for the MACHO halo fraction which is
very broad (allowing both 0 and 100\%) and which achieves its
maximum at around 30\% (see Fig.~1).
(For reference, the MACHO Collaboration bases its 50\% estimate
on their standard halo model, which is ``light'' by industry
standards, leading to their higher MACHO halo fraction.
In light of the large uncertainties the differences are not important.)
Our likelihood function for the local density of ``unknown
matter'' indicates that there is still plenty of room for
cold dark matter in our own back yard (see Fig.~2).

\begin{figure}[htb]
\epsfxsize=7.5cm
\center\leavevmode\rotate[r]{\epsfbox{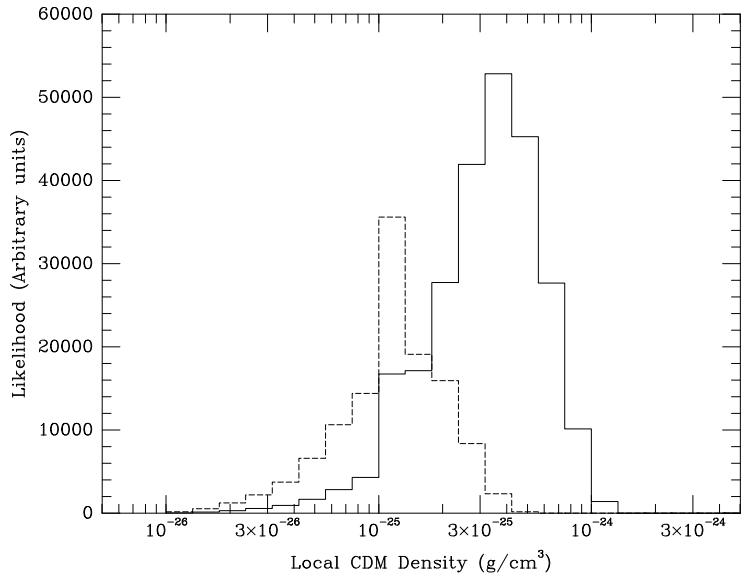}}
\caption{Monte-Carlo generated likelihood function
for the local density of cold dark matter particles, assuming
a spherical halo comprised of MACHOs and cold dark matter.  Broken curve
is the same, but with the prior that the halo MACHO
fraction is greater than 60\%.  Note, there is good reason to
believe that the halo of the Galaxy is flattened, which increases
the local halo density by about a factor of two over that
of a spherical halo \protect\cite{flat}.}
\label{fig:2}
\end{figure}

\section{Interpretation}

The MACHO Collaboration did not actually measure the MACHO fraction
of the halo!  They measured the frequency of microlensing
toward the LMC, which can be expressed as the optical
depth for microlensing -- the probability that a given LMC star
is being microlensed.  Based upon the seven events in the
two-year data set $\tau_{\rm LMC} = 2^{+1.0}_{-0.6}\times 10^{-7}$
\cite{dbatucb}.\footnote{The optical depth
$\tau = {\pi \over 4E}\sum_{i}{t_i \over \varepsilon_i}$,
where $t_i$ is the duration of event $i$, $E$ is the
exposure (e.g., in star-years), and $\varepsilon_i$ is
the efficiency of detecting an
event of duration $t_i$.}   {\it Assuming}
that MACHOs are distributed as the halo material (density decreasing
as $r^{-2}$) and that the halo MACHO fraction is constant throughout
the halo, the optical depth can be used to {\it infer} the fraction
of the halo mass contributed by MACHOs -- about 50\% according
to the MACHO Collaboration \cite{nyt,dbatucb} or 30\% according
to our analysis.

If, however, MACHOs are distributed differently, the halo MACHO fraction
can be much lower.  For example, for a spatial
density that decreases as $r^{-3}$, like the
spheroid component of the Galaxy, the inferred MACHO fraction
falls to about 10\%.\footnote{Such a model is only viable
if $\tau\la 2\times 10^{-7}$.}

Further, since the distance to the LMC is $50\kpc$ microlensing
of LMC stars cannot probe the halo beyond this.
In fact, most of the optical depth is contributed by
MACHOs that are between $10\kpc$ and $30\kpc$ from the center of
the Galaxy (see Figure in Ref.~\cite{meaning}).
It could be that the inner portion of the Galaxy has a MACHO
fraction of 50\% and that the MACHO fraction falls off beyond
this, so that something else accounts for the bulk
of the halo.  There is good evidence that the halo of the Galaxy
extends to $100\kpc$ or more.  (When the density
decreases as $r^{-2}$ the mass per interval of radial distance
is constant.)

Even more difficult to estimate is the fraction of critical
density contributed by MACHOs.  As described above,
the halo MACHO fraction is difficult to determine, and the fraction
of critical density contributed by the dark halos of spiral
galaxies is very poorly known.  This is because the full extent of dark
halos is not known; if halos are small $\Omega_{\rm HALO}$ could
be as low as 0.05; on the
other hand if halos are as large as they could be,
extending to neighboring galaxies, $\Omega_{\rm HALO}$ could be 1.0.

\section{Implications for Particle Dark Matter}

There are a number of very good reasons to believe that the bulk of matter
in the Universe exists in the form of slowly moving elementary
particles left over from the earliest moments
(cold dark matter), with the most promising cold dark matter candidates
being the axion and the neutralino.  First, the cold dark matter
scenario for structure formation is by a wide margin the most
successful, with supporting evidence coming from anisotropy measurements
of the CBR, large redshift surveys, and numerous other observations.
(By cold dark matter, we mean one of the several variants that
have been discussed -- cold dark matter plus hot dark matter,
a cosmological constant, tilted spectrum of density perturbations, unstable
tau neutrino, or very low Hubble constant \cite{appraise}.)
While the mean density of the Universe has yet to be determined,
the masses of clusters of galaxies \cite{clusters} and the peculiar
velocities of galaxies (including our own) \cite{peculiar}
indicate that $\Omega_0$ is at
least 0.2, which is larger than the upper limit to what baryons can
contribute that comes from primordial nucleosynthesis, $\Omega_B
\la 0.02h^{-2}$ \cite{cst}.  (Note, this upper limit is only close to 0.2
if the Hubble constant is very small, $h\sim 0.35$.  For
$h\sim 0.7$, $\Omega_B \la 0.04$ and virtually all cosmologists
would agree that $\Omega_0$ is greater than 0.1.)
On the theoretical side, inflation predicts a critical
density Universe, $\Omega_0 = 1.0$; current measurements of
CBR anisotropy are consistent with the inflationary prediction \cite{mwscience},
and measurements made in the next five years or so
should decisively test inflation \cite{kamio}.

It should be noted that the dark halos of spiral galaxies like
our own never provided evidence for nonbaryonic
dark matter for the simple reason that galactic halos need not
contribute more mass density than baryons can account for.  On the other hand,
galactic halos are -- and always have been --
an excellent place to search for cold dark
matter particles since it is virtually impossible to
keep cold dark matter out of galactic halos \cite{gates}.

Within the context of the cold dark matter theory the universal fraction
of matter in baryons is expected to be between 5\% and
20\%.  The baryonic content of the halo is another matter;
galactic halos probably do not provide a fair
sample of the cosmos.  At least some of the baryons in our galaxy
have undergone dissipation -- those in the disk -- and it is not
impossible that the bulk of the baryons have condensed into
the disk, in which case the
halo MACHO fraction would be smaller than the universal baryon fraction.
At the other extreme, it is possible that
most of the baryons in our galaxy have undergone only modest dissipation
and reside in the inner portion of the halo, in which case the
baryon fraction there could be 50\% or so.  Estimates
for the plausible MACHO fraction of the inner halo range from
0\% to 50\% \cite{gates}.

\section{Finale}

The MACHO, EROS, and OGLE microlensing experiments have monitored
millions of stars over the past few years looking for the proverbial
needle in the haystack.  And they have found it -- making believers of
the many nay-sayers who doubted that microlensing could be detected
against the background of known variable stars and unknown things
that go bump in the night.  The detection of microlensing represents
a significant scientific achievement as well as an important new probe
of the Galaxy.

It is still too early to make definitive statements
concerning the composition of the Galactic halo.  However,
strong evidence has been presented for the
existence of massive, dark objects within $30\kpc$ of the
center of the Galaxy.  The most plausible explanation
is dark stars (e.g., white dwarfs or neutron stars of mass
$\sim 0.3M_\odot$).

The galactic baryonic dark matter believed to
exist on the basis of big-bang nucleosynthesis may well
have been identified and one of the two dark
matter problems may have been solved!  That would be
quite an achievement.
(Big-bang nucleosynthesis provides a lower limit
to the baryon density, $\Omega_B \ga
0.01h^{-2}$, that exceeds the contribution of luminous matter,
$\Omega_{\rm LUM} \simeq 0.003h^{-1}$.)

Galactic microlensing has not diminished the case for
a Universe comprised primarily of cold dark matter.  In fact,
the inferred halo MACHO fraction
suggests that much of the nearby dark matter
in the Galaxy is still of undetermined composition (see Fig.~2),
with cold dark matter particles leading the list of
possibilities.  This, together with
the cosmic motivations for cold dark matter, should continue
to warm the hearts of WIMP chasers and justify their heroic
efforts to detect particle dark matter in our back yard.

\paragraph{Acknowledgments.}
This work was supported by the DoE (at Chicago and Fermilab) and by the NASA
(at Fermilab by grant NAG 5-2788).

\end{document}